\documentclass[12pt]{article}

\usepackage{physics} 
\usepackage{enumerate} 
\usepackage{pgfplots}
\usepackage{pgfplotstable}
\usepackage{tikz,pgfplots}
\usepackage{amsmath}  
\usepackage{wasysym} 
\usepackage{amssymb}
\usepackage{amsthm}
\usepackage{graphicx}
\usepackage{enumerate}
\usepackage{epsfig}
\usepackage{graphicx}
\usepackage{graphics}
\usepackage{float}
\usepackage{subfigure}
\usepackage{multirow}
\usepackage{lineno}
\usepackage{fullpage}
\usepackage[normalem]{ulem} 
\usepackage{makeidx}
\usepackage{xspace}
\usepackage{wrapfig}
\usepackage{relsize}

\usepackage[cmtip,all]{xy}
\usepackage{mathtools}
\usepackage{graphicx}
\usepackage{tikz-cd}
\usepackage[new]{old-arrows}

\newtheorem{theorem}{Theorem}[section]

\theoremstyle{definition}

\newtheorem*{example}{Example}

\theoremstyle{remark}
\newtheorem{remark}[theorem]{Remark}

\numberwithin{equation}{section}
\allowdisplaybreaks
\makeindex

\usepackage{geometry}
\pgfplotsset{compat=1.14}

\begin{document}

\title{A question on generalization of partition functions of CY 3-folds in String Theory} 
\author{Mohammad Reza Rahmati} 

\date{\today}  

\maketitle  



\begin{abstract} 
This is an expositoray article on the topological string partition function promoting an extension of the partition function of open Gromov-Witten theory of CY 3-folds defined by the trace of vertex operators. We also give a brief survey of their connection to the theory of Hilbert scheme of points on surface. Specifically; we apply infinitely many Casimir operators twisted to the vertex operator computing the amplitude. The case of finite number of twists has been well discussed in the mathematics and Physics literature. 
\end{abstract}

\section{Introduction}
	 
The vertex operators provide a representation theoretic and the combinatorial framework for the generating series in string theory. To a CY two generating series can be associated; one the topological vertex partition function and the other, the Guage index generating series that can be otained as the elliptic genus of certain Hilbert scheme of points on surface. The topological vertex partition function is obtained from a framed gluing of topological vertex associated to different $\mathbb{C}^3$ patches of the CY 3-fold. The partition function has various interpretations which are some how related to each other . One can mention; BPS content of the Guage actions, Fock space formalism of Hilbert scheme of points, Knot invariants, GW correlation functions, Generating series of quiver Guage theory, etc. 

The vertex operators act on the Fock space of infinite dimension and their trace produces generating series which are crucial in quantum Physics. The vertex operator provide a powerful tool to compute string theory amplitudes. We follow a computation of the character of the infinite wedge representation (Fock space) \cite{BO} where a product formula is established for the character. Bloch-Okounkov prove that the associated character is quasimodular. 

In various cases the partition function can be formulated through the Fock space structure on $\bigoplus_n H(n)$ where $H(n)$ stands for the cohomology of the Hilbert scheme of n-points on $\mathbb{C}^2$, \cite{Q}. The generating series appears as a correlation functions involving chern characters of the Hilbert scheme associated to cohomology classes on the CY 3-fold $X$. The GW-theory of CY 3-folds can be explained by the $\chi_y$-genus of the Hilbert scheme of points on surface. 

In general the successive application of commutation rule of the boson fields interacting to the vertex operator produces product factors [see Appendix in \cite{HI}]. The product formulas for the string partition functions are specially of interests. The GupaKumar-Vafa invariants can be obtained from these product formulas. 

We consider the partition functions as characters of representations and specifically as trace of vertex operators, twisted by one or more Casimir operators. We give a natural  extension of the trace to the case when infinitely many Casimirs. We ask of a representation and Physics theoretic interpretations for the trace of the extended vertex operator. 

\vspace{0.3cm}    
 
\section{Hilbert Scheme of points on Surface}

\vspace{0.3cm} 

We briefly follow \cite{Q} on basics of Hilbert scheme of points on surface. Our purpose is to present the Fock space stricture on the cohomology of Hlbert scheme of points on surface, as an alternative way to produce partition functions of CY 3-folds. 
Fix a quiver $(I,H)$, with vertex set $I$ and edges $H$. We shall consider the quiver variety 

\begin{equation}
M_{\zeta}(v,w)=\mu^{-1}(\zeta)^{ss}//GL_v \mathbb{C}
\end{equation} 

\noindent 
where $v \in \mathbb{Z}_+^I, w \in \mathbb{Z}_{\geq 0}^I$ are dimension and framing vectors and 

\begin{equation} 
\zeta=(\zeta_i) \in \mathbb{C}^I \longleftrightarrow \bigoplus_{i \in I} \zeta_i Id_{V_i} \in Z(\mathfrak{gl}_v)\end{equation}

\noindent 
$(GL_v=\prod_i GL(V_i))$, and $Z(.)$ is the center. The map $\mu$ is defined as 

\begin{equation}
\begin{aligned}
\mu:& M(v,w) \longrightarrow \mathfrak{gl}_v\\
\mu(B,a,b)&=\left ( \sum_{in(h)=i}\epsilon(h)B_hB_{\bar{h}}+a_ib_i\right )_{i \in I} \in \oplus_i \mathfrak{gl}(V_i)=\mathfrak{gl}_v
\end{aligned}
\end{equation}

\noindent 
where 

\begin{equation}
M(v,w)=\bigoplus_{h \in H}Hom(V_{out(h)},V_{in(h)}) \bigoplus \left ( Hom_{i \in H}(W_i,V_i) \oplus Hom(V_i,W_i) \right )
\end{equation}

\noindent
The variety $M_{\zeta}(v,w)$ can be interpreted as the moduli of representations of the quiver $(I,H)$. If $A=(a_{ij})$ be the adjacency matrix of the quiver $(I,H)$, then $C=2Id-A$ is a Cartan matrix, and 

\begin{equation} 
R_+=\{\beta=(\beta_i) \in \mathbb{Z}_{\geq 0}^I|\beta^TC\beta \leq 2\}
\end{equation}

\noindent 
is a set of positive roots of a Lie algebra. Lets for simplicity $(I,H)$ be a quiver of $\tilde{A}_{l-1}$-type with cyclic orientation, That is the associated lie algebra is 

\begin{equation}
\widetilde{\mathfrak{sl}}_l=\mathfrak{sl}_l[t,t^{-1}] \oplus \mathbb{C}c \oplus \mathbb{C}d
\end{equation}

\noindent 
Set 

\begin{equation}
\begin{aligned}
w_0&=(1,0,...,0) \in \mathbb{Z}^I,\\ \zeta_0&=(0, \zeta_{\mathbb{R}}), \\ \zeta_{\mathbb{R}} &\in \{\eta=(\eta_i) \in \mathbb{R}^I|\eta_i <0, \forall i\}\end{aligned}
\end{equation}

\noindent 
Let $e_i$ be the $i$-th coordinate vector of $\mathbb{Z}^I$. Define

\begin{equation}
\begin{aligned}
L_i(v)=\{(\varrho_1,\varrho_2)|\varrho_1 \subset \varrho_2,\ \ \varrho_1 \in M_{\zeta_0}(v,w_0),\ \varrho_2 \in M_{\zeta_0}(v+e_i,w_0)\}\\ \stackrel{Lagrangian}{\subset} M_{\zeta_0}(v,w_0) \times M_{\zeta_0}(v+e_i,w_0)
\end{aligned}
\end{equation}

\noindent
Consider the two projections $p_1,p_2$ from $M_{\zeta_0}(v,w_0) \times M_{\zeta_0}(v+e_i,w_0)$ onto the first and second factors. For $\alpha \in H^*(M_{\zeta_0}(v,w_0))$ and $\beta \in  H^*(M_{\zeta_0}(v+e_i,w_0))$ we can define a representation 

\begin{equation}
\begin{aligned}
e_i \longmapsto E_i:\left ( \beta \mapsto (-1)^{v_{i-1}+v_i}p_{1*}(p_2^*\beta L_i(v)) \right ), \\ 
f_i \longmapsto F_i:\left ( \alpha \mapsto (-1)^{v_{i+1}+v_i}p_{2*}(p_1^*\alpha L_i(v))\right )
\end{aligned}
\end{equation}
  
\noindent
It is possible to choose $\zeta_0$ such that 

\begin{equation}
M_{\zeta_0}(v,w_0)\cong X_0^{[n]}
\end{equation}

\noindent 
Then one gets naturally the action on $X_0^{[n]}$. Set 

\begin{equation}
H_X^{n,i}=H^i(X^{[n]}, \mathbb{C}),\ H_X^n=\bigoplus_{i=0}^{4n}H_X^{n,i}, \ H_X= \bigoplus_n H_X^n
\end{equation}

\noindent 
and 

\begin{equation}
|0 \rangle =1 \in H^0(X^{[0]})=\mathbb{C}\end{equation}

\noindent 
Define the cycles 

\begin{equation}
Q^{[m+n,m]}=\{(\xi,x,\eta)\in X^{[m+n]}\times X \times X^{[m]}|\xi \supset \eta , Supp(I_{\eta}/I_{\xi})=\{x\}\}
\end{equation}

\noindent
where $m \geq 0, Q^{[m,m]}=0$. One has $\dim Q^{[m+n,m]}=2m+n+1$. If $\alpha \in H^*(X)$ , for each $n \geq 0$ define 
the two creation and annihilation correspondences 

\begin{equation}
\begin{aligned} 
\mathfrak{a}_{-n}(\alpha):a \mapsto p_{1*}(Q^{[m+n,m]}.p^*\alpha . p_2^*a) \\
\mathfrak{a}_{n}(\alpha):b \mapsto p_{2*}(Q^{[m+n,m]}.p^*\alpha . p_1^*b) 
\end{aligned}
\end{equation}

\noindent
For $n \ne 0$, the correspondences $\mathfrak{a}_{-n}(\alpha)$ have bidegree $(n,2n-2+|\alpha|)$. We have $\mathfrak{a}_{-n}(\alpha)^{\dagger} =(-1)^n \mathfrak{a}_{n}(\alpha)$ with respect to the cup product. They satisfy the commutaion relation

\begin{equation}
[\mathfrak{a}_{m}(\alpha) ,\mathfrak{a}_{n}(\alpha)]=-m\delta_{m,-n}(\alpha,\beta)
\end{equation}

\noindent
of Heisenberg type. $H_X$ is an irreducible module over Heisenberg algebra generated by $\mathfrak{a}_n(\alpha)$ with highest weight $|0 \rangle=1 \in H^0(X^{[0]})$. It is linearly generated by 

\begin{equation}
\mathfrak{a}_{-n_1}(\alpha_1)...\mathfrak{a}_{-n_k}(\alpha_k)|0\rangle, \qquad n_1,...,n_k >0,\ \alpha_1,...,\alpha_n \in H^*(X)
\end{equation}

\noindent
One can consider the above elements factored by the image of the map $\tau_{k*}:H^*(X) \to H^*(X^k)$ induced from the diagonal embedding $\tau_k:X \to X^k$. We set 

\begin{equation}
\mathfrak{a}_{n_1}...\mathfrak{a}_{n_k}(\tau_{k*}\alpha)=\sum \mathfrak{a}_{n_1}(\alpha_{j,1})...\mathfrak{a}_{n_k}(\alpha_{j,k}), \qquad \tau_{k*}\alpha=\sum_j \alpha_{j,1}\otimes ...\otimes \alpha_{j,n}
\end{equation} 

\noindent
which is well defined cf. loc. cit. We define the normal ordering product

\begin{equation}
:\mathfrak{a}_{m_1}\mathfrak{a}_{m_2}:=\begin{cases}\mathfrak{a}_{m_1}\mathfrak{a}_{m_2}, \qquad m_1 \leq m_2 \\ \mathfrak{a}_{m_1}\mathfrak{a}_{m_2}, \qquad m_1 \geq m_2
\end{cases}
\end{equation}

\noindent 
Set 

\begin{equation}
\mathfrak{L}_n=-1/2\sum_{m \in \mathbb{Z}}:\mathfrak{a}_m\mathfrak{a}_{n-m}:
\end{equation} 

\noindent 
These operators satisfy the Virasoro relations, and also act on the former operators as shifts

\begin{equation}
\begin{aligned}
[\mathfrak{L}_m(\alpha),\mathfrak{L}_n(\beta)]&=(m-n)\mathfrak{L}_{m+n}(\alpha \beta)+\frac{m^3-m}{12}\delta_{m,-n}(\int_X e_X \alpha \beta )Id_{H_X}\\
[\mathfrak{L}_m(\alpha),\mathfrak{a}_n(\beta)]&=-n\mathfrak{a}_{m+n}(\alpha \beta)
\end{aligned}
\end{equation}

\noindent
where $e_X$ is the Euler class. 

\noindent 
Let $L$ be a line bundle on $X$. Define a virtual vector bundle $E_L$ on $X^{[k]}\times X^{[l]}$ of rank $k+l$ by 

\begin{equation}
E_L|_{(\xi,\eta)}=\chi(\mathcal{O},L)-\chi(I_{\eta},I_{\xi} \otimes L)
\end{equation}

\noindent
There is an Ext operator of Carlson 

\begin{equation}
\begin{aligned}
W(L,z):H_X[[z,z^{-1}]] \to H_Z[[z,z^{-1}]]\\
\langle W(L,z)\eta,\xi \rangle =\int_{X^{[k]}\times X^{[l]}}(\eta \otimes \xi)c_{k+l}(E_L)
\end{aligned}, \qquad \eta \in H^*(X^{[k]}), \xi \in H^*(X^{[l]})
\end{equation}

\noindent 
Define the vertex operators $\Gamma_{\pm}(L,z)=\exp(\sum_n\frac{z^{\mp n}}{n}\mathfrak{a}_{\pm n}(L))$. They satisfy the commutation relation

\begin{equation}
\Gamma_+(L).\Gamma_-(L')=(1-\frac{y}{x})^{\langle L,L'\rangle} \Gamma_-(L')\Gamma_+(L)
\end{equation}

\noindent
We can write the Carlson Ext operator in terms of vertex operators 

\begin{equation}
W(L,z)=\Gamma_-(L-K_X,z)\Gamma_+(-L,z)
\end{equation}

\noindent
where $K_X$ is the canonical class. 

Consider 

\begin{equation}
Z_n=\{(\xi,x) \subset X^{[n]}\times X| x \in Supp(\xi)\}
\end{equation}

\noindent
with two projections $p_1, p_2$. For $\gamma \in H^*(X)$ set 

\begin{equation}
G(\gamma)_n=p_{1*}\left (ch(\mathcal{O}_{Z_n})\ p_2^*td(X)\ p_2^*\gamma \right ) \in H^*(X^{[n]}), \qquad n>0
\end{equation}

\noindent
and write 

\begin{equation}
G(\gamma)_n=\oplus_iG_i(\gamma)_n, \qquad  G_i(\gamma)_n \in H^{s+2i}(X^{[n]})
\end{equation}

\noindent 
For a sheaf $F$ on $X$ define 

\begin{equation}
F^{[n]}=p_{1*}p_2^*F
\end{equation}

\noindent
This operation reduces to $K$-groups. Then one can show that

\begin{equation}
\sum_nc(L^{[n]})=\exp(\sum_{r \geq 1}\frac{(-1)^{r-1}}{r}\mathfrak{a}_{-r}(c(L)))|0 \rangle
\end{equation}

\noindent
By setting $c_{\hslash}(L^{[n]}=\sum_ic_i(L^{[n]})\hslash^i$, We can write this identity as a generating series 

\begin{equation}
\sum_nc_{\hslash}(L^{[n]}w^n)=\exp(\sum_{r \geq 1}\frac{(-1)^{r-1}}{r}\mathfrak{a}_{-r}(c_{\hslash}(L))w^r)|0 \rangle=c_{\hslash}(L)\exp(\mathfrak{a}_{-1}(1_X)w)|0 \rangle
\end{equation}

\noindent
It satisfies the differential equation

\begin{equation}
\frac{d}{dw}Z=C_{\hslash}(L)Z, \qquad C_{\hslash}(L)=c(L)\mathfrak{a}_{-1}(1_X)c(L)^{-1}
\end{equation}

\noindent
Let $L$ be a line bundle on $X$ and $ch_k(L^{[n]})$ be the $k$-th chern character of $L^{[n]}$. Define the generating series

\begin{equation}
\begin{aligned}
\langle ch_{k_1}^{L_1}...ch_{k_n}^{L_n} \rangle :&=\sum_{n \geq 0}q^n\left ( \int_{X^{[n]}}ch_{k_1}({L_1}^{[n_1]})...ch_{k_n}({L_n}^{[n_k]})c(T_{X^{[n]}}) \right )\\
\langle ch_{k_1}^{L_1}...ch_{k_n}^{L_n} \rangle&=\frac{\langle ch_{k_1}^{L_1}...ch_{k_n}^{L_n}\rangle}{\langle 1 \rangle }   \\
\langle 1 \rangle =& (q:q)_{\infty}^{-\chi(X)}, \qquad (a:q)_n=\prod_{i=0}^n(1-aq^i)
\end{aligned}
\end{equation}

\noindent
These correlation functions are conjecturally multiple $q$-zeta values, i.e. a $q$-deformation of usual multiple zeta values. They are equal to the following generating series up to a factor. Let $\alpha_1,...,\alpha_n \in H^*(X)$,

\begin{equation}
F_{k_1...k_n}^{\alpha_1...\alpha_n}=\ \sum_nq^n \left ( \int_{X^{[n]}}\prod_{i=1}^nG_{k_i}(\alpha_i,n)c(T_{X^{[n]}}) \right )=\ Tr \left ( q^nW(\mathfrak{L},s)\prod_iL_{k_i}(\alpha_i) \right ) 
\end{equation}

\noindent
They appear as the constant coefficient in a more general generating series, cf. \cite{Q}. 

\vspace{0.3cm} 

\section{Topological String partition functions of CY 3-folds}

\vspace{0.3cm}

The materials of this section are well known. The references are \cite{BHIR, KKL, HIKo, KLa, HIVa, IKS1, AKMV, IKa1, LVa, KKK, DMP, OVa, Nak, GIKV, IKV, IK, INOV, HI}. We explain the topological string partition function of CY 3-folds and the method to write it in terms of topological vertex. We employ the concept of toric varieties and their fan diagrams. Specifically we shall consider the web diagrams of CY 3-folds on the plane. 

Let $T=(\mathbb{C}^{\times})^p$ be the $p$-dimensional torus. It is the complexification of $U(1)^p$. Assume $\Sigma$ be a collection of cones called a fan in $N_{\mathbb{R}}=\mathbb{R}^m$ and $M=N^{\vee}$ the dual. Associated to a fan $\Sigma$ we can define a variety 

\begin{equation} 
M_{\Sigma}=\frac{\mathbb{C}^m \setminus Z(\Sigma)}{T_0}
\end{equation}

\noindent 
where $T_0$ is a product of torus with a finite abelian group. The equivalence classes turns out to be 

\begin{equation} 
(z_1,...,z_m) \sim (\lambda^{Q_a^1}z_1,...,\lambda^{Q_a^m}z_m), \lambda \in \mathbb{C}^{\times}
\end{equation}

\noindent 
and the relation 

\begin{equation}
\sum Q_a^iv_i=0, \qquad \text{fan condition}
\end{equation}

\noindent
where $v_i$ are generators of $\Sigma$ (called fan condition). If $D_i$ be the divisor defined by $z_i=0$, the canonical bundle of $M_{\Sigma}$ can be written as $K_M=\mathcal{O}(-\sum D_i)$. It is trivial if $\Sigma D_i \sim 0$. This condition can be interpreted as the existence of a lattice point $m \in (\mathbb{R}^m)^{\vee}$ such that $\langle v_i,m \rangle =1$ for all $i$. Thus $M_{\Sigma}$ is CY iff the vectors $v_i$ all lie in a hyperplane. It follows that 

\begin{equation}
\sum_i Q_a^i=0, \qquad \text{CY condition}
\end{equation}

\noindent
Thus a toric CY is never compact. 

We will deal with CY 3-folds. In this case the 3-dimensional fan is projected on a 2-dimensional graph, called the web diagram, and $T_0=(\mathbb{C}^{\times})^{m-3} \times \text{finite abelian gp}$. There is a moment map

\begin{equation}
\mu:M_{\Sigma} \to \Delta_M
\end{equation} 

\noindent 
where $\Delta_M$ is the polytope of $M$, i.e is dual to the fan $\sigma$. The moment map is given by $m-3$ maps $\mu_a:\mathbb{C}^m \to \mathbb{C}, \ a=1,2,...,m-3$, and is described by the equations 

\begin{equation}
\sum_{i=1}^mQ_a^i|z_i|^2=Re(t_a), \qquad t_a \in \mathbb{C}
\end{equation}

\noindent
called Witten $D$-term equations. We have an action of $U(1)^{m-3}$ on coordinates $z_j$

\begin{equation}
z_j \longmapsto \exp(iQ_a^i \alpha_a)z_j , \qquad a=1,2,...,m-3
\end{equation}

\noindent
It turns out that 

\begin{equation}
M_{\Sigma}=\frac{\bigcap_{i=1}^{m-3}\mu^{-1}(Re(t_a))}{T_0}
\end{equation}

\noindent
is a CY 3-fold and $t_a$ are complexified Kahler parameters. We look at CY 3-folds as $T^3=(S^1)^3$-fibrations over 3-dimensional base with corners. Introduce new variables 

\begin{equation}
(p_1, \theta_1),...,(p_k,\theta_k), \qquad  p_i=|z_i|^2, \ z_i=|z_i|e^{i\theta_i}
\end{equation} 

\noindent 
The 3-fold is then parametrized by the coordinates $p_i$ and $\theta_i$, where the latter describes $T^3$. The CY 3-fold is constructed by gluing many open $\mathbb{C}^3$-patches to each other along edges. The web diagram of CY 3-folds records the information of a framing when gluing $\mathbb{C}^3$-patches. The frame is settled by choosing a vector $f_i$ perpendicular to the vectors $v_i$ of the fan. The choice of framing affects the partition function of the CY, however here we assume the frame is chosen to be standard, i.e $f_i \wedge v_i=1$ for all $i$.\\

\noindent 
\textbf{Fermionic model:} \cite{AKMV} To each $\mathbb{C}^3$-patch of the CY variety we associate a partition function called topological vertex $C_{R_1,R_2,R_3}$ which can be written in the following two equivalent forms 

\begin{equation}
\sum_{R_1,R_2,R_3}C_{R_1,R_2,R_3}^{f_1,f_2,f_3}\prod_{i=1}^3{Tr_R}_iV_i=\sum_{
\overrightarrow{k}(1),\overrightarrow{k}(2),\overrightarrow{k}(3)}
C_{\overrightarrow{k}(1),\overrightarrow{k}(2),\overrightarrow{k}(3)}^{f_1,f_2,f_3}
\prod_{i=1}^3\frac{1}{n_{\overrightarrow{k}(i)}}Tr_{\overrightarrow{k}(i)}V_i
\end{equation}

\noindent
where $Tr_RV$ denotes the trace of $V$ as appears in the representation $R$, and $\chi_R(C(\overrightarrow{k}))$ is the character of the symmetric group calculated at the conjugacy class $C(\overrightarrow{k})$. The sum in the left hand side runs over all the representations $R_1,R_2,R_3$ of the symmetric group $S_N$ for all $N$. The trace of representations of the symmetric groups can also be stated via the conjugacy classes and we simply have 

\begin{equation}
Tr_{\overrightarrow{k}}V=\sum_R\chi_R(C(\overrightarrow{k}))Tr_RV
\end{equation}

\noindent 
The two sums are related by the following identity

\begin{equation}
C_{R_1,R_2,R_3}^{f_1,f_2,f_3}=\sum_{
\overrightarrow{k}(1),\overrightarrow{k}(2),\overrightarrow{k}(3)}
C_{\overrightarrow{k}(1),\overrightarrow{k}(2),\overrightarrow{k}(3)}^{f_1,f_2,f_3}
\prod_{i=1}^3\frac{{\chi_R}_i\left ( C(\overrightarrow{k}(i)) \right )}{n_{\overrightarrow{k}(i)}}
\end{equation}

\noindent
The $V_i$ are holonomy variables, given by the determinant of the Wilson line $V_i =P \int_{\Gamma_i}A_i$. In fact one notes that in counting the holomorphic maps from curves of different genus to a CY 3-fold, these curves must go to the vertices or edges of the web diagram, and to illustrate them, they wrap arround the edges with different holonomies. The change of the framing affect as

\begin{equation}
C_{R_1,R_2,R_3}^{f_1-n_1v_1,f_2-n_2v_2,f_3-n_3v_3}=(-1)^{\sum n_il(R_i)}q^{\sum_in_ik(R_i)/2}C_{R_1,R_2,R_3}^{f_1,f_2,f_3}
\end{equation}

\noindent
[See \cite{AKMV} for the definition of indices $l(R_i), k(R_i)$ and $n_{\overrightarrow{k}(i)}$]. The topological vertex $C_{R_1,R_2,R_3}$ is invariant under circle change of the representations $R_i$, i.e they have cyclic symmetry. We have the rule 

\begin{equation}
v_1 \to -v_1 \leadsto C_{R_1,R_2,R_3} \to (-1)^{l(R_1)}C_{R_1^t,R_2,R_3}
\end{equation}

\noindent 
In the gluing of two graphs $\gamma_1$ and $\Gamma_2$ with partition functions $Z(\Gamma_1)$ and $Z_{\Gamma_2}$ we produce terms as

\begin{equation}
\sum_QZ(\Gamma_1)(-1)^{l(Q}e^{l(Q)t}Z(\Gamma_2)_{Q^t}\qquad \longleftrightarrow \qquad \sum_{\overrightarrow{k}}Z(\Gamma_1)\frac{\exp(-l(\overrightarrow{k})t)}{\prod_jk_j!j^{k_j}} Z(\Gamma_2)
\end{equation}

\noindent 
The rules to write the partition function in terms of topological vertex are as follows,

\begin{itemize}
\item[(1)] The edges of the graph are labeled by integral vectors $v_i$. To each edge associate a representation $R_i$.
\item[(2)] For smooth CY the graph can be divided into trivalent vertices corresponding to $\mathbb{C}^3$-patches.
\item[(3)] To each vertex there associates an ordered triple $(v_i,v_j,v_k)$ by reading counterlockwise.
\item[(4)] If all the edges are incoming associate $C_{R_i,R_j,R_k}$ to $(v_i,v_j,v_k)$, otherwise replace the corresponding representation by its transpose times $(-1)^{l(R)}$. 
\item[(5)] If the vertex $(v_i,v_j,v_k)$ shares the $i$-th edge with the vertex $(v_i,v_{j'},v_{k'})$, we glue the amplitudes by summing over the representations on the $i$-th edge

\begin{equation}
\sum_{R_i}C_{R_jR_kR_i}\left ( e^{-l(R_i)t_i}(-1)^{(n_i+1)l(R_i)}q^{-n_ik(R_i)/2}\right ) C_{R_i^tR_{j'}R_{k'}}
\end{equation}

\noindent
where $n_i=|v_k \wedge v_{k'}|$.
\item[(6)] The length of edge can be read from the Witten D-term equation on the Kahler moduli of $X$. The edges of the graph $\Gamma$ are straight lines on the plane with rational slope. To the $i$-th edge in the $(p_i,q_i)$-direction of length $x_i$ we associate a Kahler parameter $t_i=x_i/\sqrt{p^2+q^2}$. 
\item[(7)] For non-compact edges the corresponding representation is trivial, denoted $R=0, \emptyset, \bullet$.
\item[(8)] If there are more than one $D$-branes say $n$ on the edge, we have contributions of ${n \choose 2}$ open strings stretching between $D$-branes. The effect of integrating out these strings is $\exp(-\sum\frac{1}{m}trU_1^mU_2^m)=\sum_R(-1)^{l(R)}tr_RU_1tr_RU_2$. It produces contributions of the form

\begin{equation}
\sum_{R_i,Q_{a,i}^L,Q_{a,i}^R}C_{R_j,R_k,R_i \otimes_{a=1}^nQ_{a,i}^L} \left ( (-1)^{s(i)}e^{-L(i)}q^{f(i)}\right ) C_{R_i^t\otimes_{a=1}^nQ_{i,a}^R,R_j',R_k'}\prod_{a=1}^nTr_{Q_{a,i}^L}V_aTr_{Q_{a,i}^R}V_a^{-1}.
\end{equation}

\noindent
where the exponents in the middle term is read from the Young diagram of the representations $R_l, R_l', Q_{a,i}^L, Q_{a,i}^R$, cf. \cite{AKMV}.
\item[(9)] The topological vertex is given by 

\begin{equation}
C_{\lambda\mu\nu}=q^{|k(\mu)/2}S_{\nu^t}(q^{-\rho})\sum_{\eta}S_{\lambda^t/\eta}(q^{-v-\rho})S_{\mu/\eta}(q^{-\nu^t-\rho})
\end{equation}

\noindent
where $q^{\nu-\rho}=\{q^{-\nu_i+(2i-1)/2}|i=1,2,...\}, \ q=e^{2\pi i\epsilon}$, and $\parallel \lambda \parallel=\sum \lambda_i^2, \ k(\mu)=|\mu|-\parallel \mu \parallel$. 
\end{itemize}   

\vspace{0.3cm}

\noindent
\textbf{Bosonic model:} \cite{AKMV} We explain another model to express the partition function of CY 3-folds involving Boson operators. We shall consider the Hilbert space $\mathcal{H}$ of states to be generated by 

\begin{equation}
|\overrightarrow{k}\rangle =\prod_j \alpha_{-j}^{k_j}|0 \rangle, \qquad \overrightarrow{k}=(k_1,k_2,...,k_m)
\end{equation}

\noindent 
is a vector involving winding numbers. On $\mathcal{H}^{\otimes 2}$ we define an element 

\begin{equation}
|P \rangle =\exp(-t\sum_j\frac{1}{n}\alpha_{-j}\alpha_j)|0 \rangle \otimes  \langle0|=\sum_{\overrightarrow{k}}e^{-l(\overrightarrow{k})t}\frac{(-1)^h}{n_{\overrightarrow{k}}}|\overrightarrow{k}\rangle \otimes \langle \overrightarrow{k}|, \qquad n_{\overrightarrow{k}}=\prod_jk_j!j^{k_j}
\end{equation}

\noindent
called propagator. The topological vertex is defined as a state in $\mathcal{H}^{\otimes 3}$. If we write the partition function $Z$ formally as 

\begin{equation}
Z=\sum_{\overrightarrow{k}_i}C_{\overrightarrow{k}_1,\overrightarrow{k}_2,\overrightarrow{k}_3}\frac{1}{\prod_in_{\overrightarrow{k}_i}}Tr_{\overrightarrow{k}_1}V_1Tr_{\overrightarrow{k}_2}V_2Tr_{\overrightarrow{k}_3}V_3
\end{equation}

\noindent 
Thinking of topological vertex $C$ as a state in $\mathcal{H}^{\otimes 3}$ we write $Z$ by 

\begin{equation}
Z=\sum_{\overrightarrow{k}_i}Tr_{\overrightarrow{k}_1}V_1Tr_{\overrightarrow{k}_2}V_2Tr_{\overrightarrow{k}_3}V_3\frac{1}{\prod_in_{\overrightarrow{k}_i}}\langle \overrightarrow{k}_1| \otimes \langle \overrightarrow{k}_2| \otimes \langle \overrightarrow{k}_3|C \rangle 
\end{equation}

\noindent 
It follows from the last formula that $C$ can be written in the form 

\begin{equation}
|C \rangle =\exp \left (\sum_{\overrightarrow{k}_i}F_{\overrightarrow{k}_1,\overrightarrow{k}_2,\overrightarrow{k}_3}(g_s)\alpha_{-\overrightarrow{k}_1}\alpha_{-\overrightarrow{k}_2}\alpha_{-\overrightarrow{k}_3}\right )| 0 \rangle_1 \otimes |0 \rangle_2 \otimes 0 \rangle_3
\end{equation}

\noindent
We may formally write $Z=\langle V_1 | \otimes \langle V_2 | \otimes \langle V_3 |C \rangle$ where 

\begin{equation}
V=\langle 0 |\exp \left ( \sum \frac{1}{n}Tr V^{\otimes n} \alpha_n \right ) 
\end{equation}

\begin{example}
\begin{itemize}
\item The partition function of the web diagram obtained by gluing three $\mathbb{C}^3$ patches along a triangle is 

\begin{equation}
Z=\sum_{R_1,R_2,R_3}(-1)^{\sum_i l(R_i)}e^{-l(R_i)t}q^{\sum_i k_{R_i}}C_{\bullet R_2R_3^t}C_{\bullet R_1R_2^t}C_{\bullet R_3R_1^t}
\end{equation}

\item The partition function of an $(M,N)$-web diagram with $M$ vertical and $N$ horizontal hexagons is written as

\begin{equation}
Z(M,N)=\sum_{\alpha}\prod_{a=1}^NQ_{b_a}^{|\alpha^a|}\sum_{\mu,\nu}(-Q_m)^{\sum_b|\mu_b|-|\nu_b|}\prod_b (-Q_b)^{|\nu_{b+1}|}\prod_aC_{\mu_a\nu_a\alpha_a}(t,q)\prod_bC_{\nu_a^t\mu_a^t\beta_a^t}(t,q)
\end{equation}

\noindent
where $\alpha^a$ is a set of $M$ partitions $\alpha_1^a,...,\alpha_M^a$, and $\beta_b=\alpha_b^{a+1}$, \cite{HI}.
\end{itemize}
\end{example}

\vspace{0.5cm}

\section{Vertex Operator formalism of Partition Functions}

\vspace{0.3cm} 

The material of this section are well known. The references are \cite{Ze, AKV1, IKS1} as well as many other available texts. We include this section as part of the task to make our terminology more concrete and understandable. We begin with the definition of the Fock space. The Fock space $\mathfrak{F}=\bigwedge^{\frac{\infty}{2}}V$ is the vector space spanned by semi-infinite wedge product of a fixed basis of the infinite dimensional vector space $V =\sum_{i \in 1/2 +\mathbb{Z}}\mathbb{C}v_i$, i.e. monomials $v_{i_1} \wedge v_{i_2} \wedge ...$ such that 

\begin{itemize}
\item $i_1 >i_2>...$
\item $i_j=i_{j-1}-1/2$ for $j \gg 0$
\end{itemize}

\noindent
We have the creation operators and and the annihilation operators

\begin{equation}
\begin{aligned}
\psi_k:&v_{i_1} \wedge v_{i_2} \wedge ... \mapsto v_k \wedge v_{i_1} \wedge v_{i_2} \wedge ...\\
\psi_k^*:&v_{i_1} \wedge v_{i_2} \wedge ... \mapsto (-1)^l v_{i_1} \wedge v_{i_2} \wedge ...\wedge \widehat{v_{i_l=k}} \wedge ...
\end{aligned}
\end{equation}

\noindent
The monomials can be parametrized by partitions 

\begin{equation}
|\lambda \rangle =v_{\lambda}=\lambda_1-1/2 \wedge \lambda_2 -3/2 \wedge ...
\end{equation}

\noindent
We can also write this using Frobenius coordinates of partitions

\begin{equation}
|\lambda \rangle =\prod_{i=1}^l\psi_{a_i}^*\psi_{b_i}|0 \rangle ,\qquad a_i=\lambda_i-i+1/2,\ b_i=\lambda_i^t-i+1/2
\end{equation}

\noindent
We can define operators 

\begin{equation}
\alpha_n=\sum_{k \in 1/2 +\mathbb{Z}} \psi_{k+n}\psi_k^*
\end{equation}

\noindent
They satisfy $[\alpha_n, \psi_k]=\psi_{k+n},\ [\alpha_n, \psi_k^*]=-\psi_{k-n}$.
The operators of the form 

\begin{equation}
\Gamma_+(x)=\exp(\sum_{n \geq 1}\frac{x^n}{n}\alpha_n), \qquad \Gamma_-(x)=\exp(\sum_{n >0} \frac{x^n}{n}\alpha_{-n})
\end{equation}

\noindent
are called vertex operators. They are adjoint with respect to the natural inner product. We have a commutation relation

\begin{equation}
\Gamma_+(x)\Gamma_-(y)=(1-xy)
\Gamma_-(y)\Gamma_+(x)
\end{equation}

\noindent
We have 

\begin{equation}
\Gamma_+(x) v_{\mu}=\sum_{\lambda \supset \mu}s_{\lambda/\mu}(x)v_{\lambda}
\end{equation} 

\noindent
Vertex operators provide powerful tools to express partitions. 

\begin{equation}
\begin{aligned}
\Gamma_+(1) |\mu \rangle =\sum_{\lambda \supset \mu}|\lambda \rangle\\
\Gamma_-(1) |\mu \rangle =\sum_{\lambda \subset \mu}|\lambda \rangle 
\end{aligned}
\end{equation}

\noindent
For example we may write the McMahon function as 

\begin{equation}
\begin{aligned}
Z&=\sum_{\text{3-dim partitions}}q^{\natural \ boxes} = \langle (\prod_{t=0}^{\infty}q^{L_0}\Gamma_+(1)) q^{L_0} (\prod_{t=-\infty}^{-1}\Gamma_-(1)q^{L_0})\rangle\\
&=\langle \prod_{n>0}\Gamma_+(q^{n-1/2})\prod_{n>0}\Gamma_-(q^{-n-1/2}) \rangle
\end{aligned}
\end{equation}

\noindent 
We may divide a 3-dimensional partitions into slices of two dimensional partitions , for instance along the diagonals or any other way this could be done. In this way the vertex operator divides into multiplication of many vertex operators of the slices,

\begin{equation}
Z(\{x_m^{\pm}\})=\langle ... \prod_{u_i< m<v_{i+1}}\Gamma_+(x_m^-)\prod_{v_i< m<u_{i}}\Gamma_-(x_m^+)...\rangle = \langle \prod_{u_0< m<u_n}\Gamma_{-\epsilon(m)}(x_m^{\epsilon(m)})  \rangle
\end{equation}

\noindent
In this way one can obtain product formulas such as 

\begin{equation}
Z(\{x_m^{\pm}\})\prod_{m_1 <m_2} (1-x_{m_1}^-x_{m_2}^+)
\end{equation}

\noindent
For example we may choose 

\begin{equation}
\begin{aligned}
\{x_m^{\pm}\}&=\{t^iq^{v_i}|i=1,2,...\}\\
\{x_m^{\pm}\}&=\{t^{j-1}q^{-v_i^t}|j=1,2,...\}
\end{aligned}
\end{equation}

\noindent
Then we get

\begin{equation}
Z_{\lambda\mu\nu}(t,q)=\langle \prod_{u_0< m<u_n}\Gamma_{-\epsilon(m)}(x_m^{\epsilon(m)}) \rangle 
\end{equation}

\noindent
The partition function can also be read by putting a wall on the distance $M$ along one of the axis. Then using commutation (7.6) ,we have expressions of the form

\begin{equation}
\begin{aligned}
Z&=\langle \prod_{0<m <\infty}\Gamma_-(x_m^+)\prod_{-M<m<0}\Gamma_+(x_m^-) \rangle  \\
&=\prod_{l_1=1}^{\infty}\prod_{l_2=1}^M(1-x_{l_1-1/2}^+x_{-l_2+1/2}^-)^{-1}(\langle \prod_{-M<m<0}\Gamma_+(x_m^-)\prod_{0<m<\infty}\Gamma_-(x_m^+) \rangle
\end{aligned}
\end{equation}

\noindent
The last factor in paranthesis is equal to $1$, and we obtain a product formula. 

\begin{example}
The refined partition function of the CY 3-fold double $\mathbb{P}^1$ (3 $\mathbb{C}^3$ patches where each of the outer two glued to the middle one along a a compact leg of length $Q_i, \ i=1, 2$) can be written as

\begin{equation}
\begin{aligned}
Z'&=\langle \prod_{0<m <L}\Gamma_-(x_m^+)\prod_{-M<m<0}\Gamma_+(x_m^-) \rangle \\
&=\sum_{\lambda}\langle 0 |\prod_{0<m <L}\Gamma_-(x_m^+)|\lambda \rangle \langle \lambda|\prod_{-M<m<0}\Gamma_+(x_m^-)|0 \rangle\\ 
&= \prod_i \prod_j \frac{(1-Q_1t^{i-1/2}q^{j-1/2})(1-Q_2t^{i-1/2}q^{j-1/2})}{(1-Q_1Q_2t^{i-1}q^{j})}
\end{aligned}
\end{equation} 

\noindent
where $\lambda$ runs over partitions in the range, and 

\begin{equation}
\begin{aligned}
\{x_m^+\}=\{t,t^2,...,t^L\}\\
\{x_m^-\}=\{1,q,q^2,...,q^{M-1}\}
\end{aligned}
\end{equation}

\noindent
$Z'=Z_{\text{McMahon}}Z_{\text{refined}}$, \cite{CIK}.
\end{example} 

\begin{remark}
The quotient $\frac{Z{\lambda\mu\nu}}{Z_{\emptyset\emptyset\emptyset}}=C_{\lambda\mu\nu}$ is called refined topological vertex. The aforementioned technics also applies to the refined version of topological vertex. 
\end{remark} 

\noindent 
To give some sense of computations we calculate the trace of a general vertex operator action on the Fock space $\mathfrak{F}$, cf. \cite{Ze}. 
We have the following formula for the trace of a vertex operator acting on $\mathfrak{F}=\bigwedge^{\frac{\infty}{2}}V$;

\begin{equation}
Tr \left ( q^{L_0} \exp ( \sum_n A_n \alpha_{-n}) \exp ( \sum_n B_n \alpha_n ) \right )= \prod_n \sum_{k}\sum_{l=0}^k\frac{n^lA_n^lB_n^l}{l!}q^{nk}{k \choose l} 
\end{equation}

\noindent
where $L_0$ is the charge operator, $q^{L_0}| \lambda \rangle =|\lambda ||\lambda \rangle$.

\noindent 
To prove the formula, denote the operator in the trace by $T$. We have the isomorphism 

\begin{equation}
\bigwedge^{\frac{\infty}{2}}V= \bigotimes_{n=1}^{\infty} \bigoplus_{k=0}^{\infty} \alpha_{-n}^k|0 \rangle
\end{equation}

\noindent
which implies

\begin{equation}
\begin{aligned}
Tr(T)&=\prod_{n=1}^{\infty} Tr \left ( T|_{\bigoplus_{k=0}^{\infty} \alpha_{-n}^k|0 \rangle} \right )\\
&= \prod_n \sum_k \langle \alpha_{-n}^k|0 \rangle \Big | q^{L_0}e^{A_n\alpha_{-n}}e^{B_n\alpha_n} \Big | |\alpha_n^k|0 \rangle \rangle\\
&= \prod_n \sum_{k,l,m}\frac{A_n^lB_n^m}{l!m!}q^{n(l-m+k)} \langle \alpha_{-n}^k|0 \rangle \Big | \alpha_{-n}^l\alpha_n^m \Big | |\alpha_n^k|0 \rangle \rangle\\
&= \prod_n \sum_{k,l}\frac{A_n^lB_n^l}{l!l!}q^{nk} \langle \alpha_{-n}^k|0 \rangle \Big | \alpha_{-n}^l\alpha_n^l \Big | |\alpha_n^k|0 \rangle \rangle\\
&= \prod_n \sum_{k}\sum_{l=0}^k\frac{n^lA_n^lB_n^l}{l!}q^{nk}{k \choose l} 
\end{aligned}
\end{equation}

\vspace{0.3cm}

\section{Generalization} 

\vspace{0.3cm} 

In this section we extend the partition function of a CY 3-fold as the trace vertex operators twisted by certain Casimir operators, and present our main result. Our motivation is a computation on the character of the infinite wedge Fock space in \cite{BO}. Let $\mathfrak{F}=\bigwedge^{\frac{\infty}{2}}V$ be the Fock space on a fixed basis of $V=\bigoplus_{j \in \mathbb{Z}} \mathbb{C}v_j$. Then the character of $\mathfrak{F}$ as a representation of $\mathfrak{gl}_{\infty}$ is given by

\begin{equation}
ch(\mathfrak{gl}_{\infty}, \mathfrak{F})= \prod_{n \geq 0} (1+q_0q_1^{n+1/2}q_2^{n+3/2)^2}....)(1+q_0^{-1}q_1^{n-1/2}q_2^{-(n-1/2)^2}....)
\end{equation}

\noindent
where $q_j=e^{2\pi i \tau_j}$. The partition function of $U(1)$ theory can be written in the form 

\begin{equation}
Z(\tau, m, \epsilon)=Tr \left ( Q_{\tau}^{L_0} \exp \left (\sum_{n \geq 1}\frac{Q_m^n-1}{n(q^{n/2}-q^{-n/2})}\alpha_n \right ) \exp \left (\sum_{n \geq 1}\frac{Q_{-m}^n-1}{n(q^{n/2}-q^{-n/2})}\alpha_{-n} \right ) \right ) 
\end{equation}

\noindent
Using the commutation relation of $\alpha_{\pm}$ it can be written as 

\begin{equation}
Z(\tau, m, \epsilon)=\prod_k(1-Q_{\tau}^k)^{-1}\prod_{i,j}\frac{(1-Q_{\tau}^kQ_m^{-1}q^{i+j-1})(1-Q_{\tau}^kQ_mq^{i+j-1})}{(1-Q_{\tau}^kq^{i+j-1})}
\end{equation}

\noindent
cf. \cite{CIK}. The partition function in (8.4) can be generalized to

\begin{equation}
Z(\tau, m, \epsilon,t)=Tr \left ( Q_{\tau}^{L_0} e^{\sum_nt_nL_n} \exp \left (\sum_{n \geq 1}\frac{Q_m^n-1}{n(q^{n/2}-q^{-n/2})}\alpha_n \right ) \exp \left (\sum_{n \geq 1}\frac{Q_{-m}^n-1}{n(q^{n/2}-q^{-n/2})}\alpha_{-n} \right ) \right ) 
\end{equation}

\noindent
In the limit $m \mapsto 0$  we obtain 

\begin{equation}
Z(\tau, m=0, \epsilon,t)=Tr \left ( Q_{\tau}^{L_0} e^{\sum_nt_nL_n} \right )
\end{equation}

\noindent 
We can write the partition function in Remark 8.2 in terms of the Gromov-Witten potentials 

\begin{equation}
Z(\tau, m, \epsilon)=\exp\left (\sum_{g \geq 0}\epsilon^{2g-2}F_g \right )
\end{equation}

\noindent
where 

\begin{equation}
e^{F_1}=\prod_k(1-Q_{\tau}^k)^{-1}\left ( \frac{(1-Q_{\tau}^k)^{2}Q_m^{-1})(1-Q_{\tau}^kQ_m)^{2}}{(1-Q_{\tau}^k)^{4}} \right )^{1/24}
\end{equation}

\noindent
One may ask if the above equation is equal to 6.8 with $q_2=q_3=....=1$.

\vspace{0.5cm}

\noindent 
\textbf{Problem:} 
Consider the following trace as a twist of the two former ones,

\begin{equation}
Tr \left ( \exp(\sum_{j\geq 0} 2 \pi i L_j) \exp ( \sum_{n > 0} A_n \alpha_{-n}) \exp ( \sum_{n > 0} B_n \alpha_n ) \right )
\end{equation}

\noindent
We pose the following questions; 

\begin{itemize}
\item How to compute the trace in terms of the former traces. 
\item How the last trace is related to the former two characters in Theorems 8.1 and 8.2. What is the representation theory interpretation of that.
\item In case that the coefficients $A_n, B_n$ are suitably chosen what is the Physical interpretation of the trace in terms of string theory partition functions.
\item Is there any product formula for the trace.
\end{itemize}

\end{document}